# Determination of Power of Groove fields belonging to the wedge regions adjacent to a convex triangular obstacle associated with Dirichlet conditions subject to axially independent EM fields

**Sanjay Kumar and Suresh K. Shukla**

P. G. Centre Department of Mathematics
Ram Dayalu Singh College, Muzaffarpur-842002, Bihar, India
Email: dr.sanjaykumar06@gmail.com; shuklasureshk@gmail.com

**Abstract:** A convex triangular obstacle forms a vital part of a periodic echellete grating. A triangular grating is characterized by three parameters like period, depth and flare angle. Knowledge of groove field is essential for precise designing of triangular corrugated structures for studying the blazing effect of propagating EM wave. In the present paper, an attempt has been made to determine the power of Groove fields belonging to a pair of groove regions adjacent to a convex triangular prism. Groove fields and their associated powers based on Dirichlet conditions on the groove surfaces have been determined. The governing Helmholtz wave equation has been solved for determining the free surface field and the groove field. Fourier-Bessel series, oblique coordinate transformations and Lommel's integral are used as tools.

1. **Introduction**

A convex triangular obstacle forms a vital part of a periodic echellete grating. During recent years [1-8] quite a good number of results have been reported pertaining to the groove field estimates and the efficiency of the said grating. The present paper deals with a smooth convex triangular prismatic obstacle $K$ having an open base, a flare angle β, the groove depth '$h$' and the grating period '$d$' (Fig. 1). The bounding faces $\partial K$ of the obstacle $K$ and its adjacent wedge surfaces (Fig. 2) are subjected to reflection, transmission and grazing due to an axially independent EM wave. EM field intensities $\mathbf{F} = (\mathbf{H} \vee \mathbf{E})$ are derived from the governing Maxwell's equations

$$\nabla \times \mathbf{H} = \mathbf{J} = \sigma \mathbf{E} + \varepsilon \frac{\partial \mathbf{E}}{\partial t}, \qquad \nabla \times \mathbf{E} = -\frac{\partial \mathbf{B}}{\partial t} = -\mu \frac{\partial \mathbf{H}}{\partial t}$$

$$\text{and} \quad \nabla^2 \mathbf{F} = \mu \left( \sigma \frac{\partial \mathbf{F}}{\partial t} + \varepsilon \frac{\partial^2 \mathbf{F}}{\partial t^2} \right), \tag{1}$$

where $H$ and $E$ stand for the magnetic and the electric intensity vectors. The physical elements σ, $\varepsilon$ and μ being associated with the medium for the EM wave and the constituent material for $\partial K$, stand for conductivity, permittivity and the permeability respectively. Maxwell's equations (1) have been encountered subject to prescribed initial boundary conditions of the EM field on $\partial K$. The concerning boundary value problems happen to be associated with the Dirichlet's conditions $F|_{\partial K} = f$ initially (t = 0) on $\partial K$. Dirichlet's problem is an example of well posed boundary value problem as observed earlier [9-11]. An axially independent field intensity satisfies the condition $\partial \mathbf{F}/\partial x_3 = 0$, which leads to the independence of $\mathbf{F}$ relative to the directions parallel to the edges $OO'$, $AA'$ and $BB'$ of the model '$K$'. A wave is said to be cylindrically polarized whenever the constituent particles of the wave are constrained to vibrate on the surface of a cylinder. As such, a cylindrical wave function happens to exist as a solution of the Maxwell's equation (1) subject to cylindrical coordinate transformation $x_1 = \rho \cos\phi$, $x_2 = \rho \sin\phi$, $x_3 = x_3$. Method of separation of variables has been utilized for arriving at the solution in the form of the Fourier-Bessel series [12]

$$\mathbf{F} = \mathbf{F}(\rho, \phi, t) = \sum_{i \in J^+} \mathbf{A}_i\, J_\eta(\rho k_i) \exp\left\{ \left( j\omega t - \frac{\sigma t}{2\varepsilon} \right) + j\eta\phi \right\} \tag{2}$$

where $\eta \geq 1$ and $k_i$ is the $i^{th}$ wave number in a certain frequency range associated with interacting EM waves.

The unknown coefficients '$A_i$' happen to satisfy two pairs of dual-Bessel series relations in the wedge regions $R_i$ ($i = 1, 2$). Oblique coordinate transformations [12] being associated with the geometry of '$K$', have been found to be of great value for evaluating the coefficients '$A_i$'. It is essential to add that '$A_i$' may be evaluated by means of the integral formula

$$\frac{1}{2} A_i d^2 J_{\eta+1}^2 (dk_i) = \int_{\rho=0}^{d} \rho J_\eta (k_i \rho) f(\rho) d\rho \qquad (3)$$

where $k_i$ is the $i^{th}$ positive zero of $J_\eta(dk) = 0$, and the function $f(\rho)$ is prescribed on $\partial K$ (Dirichlet condition). Hence the EM field $\mathbf{F}$ is completely determined by substituting the values of '$A_i$' in the Fourier-Bessel series (2).

Finally, the expressions of $\mathbf{F}$ have been used for computing the current density. The paper is concluded by arriving at the expressions of the powers of the groove fields.

2. **Formulation of the problem**

Consider the Maxwell's equation

$$\nabla^2 \mathbf{F} = \frac{\partial^2 F}{\partial x_1^2} + \frac{\partial^2 F}{\partial x_2^2} + \frac{\partial^2 F}{\partial x_3^2} = \mu \left( \sigma \frac{\partial \mathbf{F}}{\partial t} + \varepsilon \frac{\partial^2 \mathbf{F}}{\partial t^2} \right) \qquad (4)$$

where $\mathbf{F} = \mathbf{H} \vee \mathbf{E} = \mathbf{F}(x_1, x_2, x_3, t)$ stands for vector field intensity. Transforming (4) by using cylindrical coordinates $x_1 = \rho \cos\phi$, $x_2 = \rho \sin\phi$, $x_3 = x_3$ subject to axially independent condition $\frac{\partial \mathbf{F}}{\partial x_3} = 0$, one can arrive at the equation

$$\frac{\partial^2 \mathbf{F}}{\partial \rho^2} + \frac{1}{\rho} \frac{\partial \mathbf{F}}{\partial \rho} + \frac{1}{\rho^2} \frac{\partial^2 \mathbf{F}}{\partial \phi^2} = \mu \left( \sigma \frac{\partial \mathbf{F}}{\partial t} + \varepsilon \frac{\partial^2 \mathbf{F}}{\partial t^2} \right) \qquad (5)$$

Now, using the variable separable method, one can arrive at the solution of the equation (5) in the form

$$\mathbf{F}(\rho, \phi, t) = F_1(\rho) F_2(\phi) F_3(t) \qquad (6)$$

where $F_1$, $F_2$ and $F_3$ satisfy the ordinary differential equations

$$\rho^2 F_1'' + \rho F_1' + (k^2 \rho^2 - \eta^2) F_1 = 0 \qquad (7)$$

$$F_2'' + \eta^2 F_2 = 0 \qquad (8)$$

and
$$\mu \varepsilon F_3'' + \mu\sigma F_3'' + k^2 F_3 = 0 \tag{9}$$

The equations (7), (8) and (9) furnish the solutions

$$F_1(\rho) = J_\eta(k\rho), \quad F_2(\phi) = A e^{\eta\phi j} \text{ and } F_3 = B\exp\left[-t\left(\frac{\sigma}{2\varepsilon} - j\omega\right)\right], \quad j = \sqrt{-1} \tag{10}$$

$$\mu^2\sigma^2 - 4\mu\varepsilon k^2 = -4\omega^2\mu^2\varepsilon^2 \tag{11}$$

where $J_\eta(\rho k)$ is the Bessel function of the first kind of order $\eta$, and A and B are arbitrary constants.

**Fourier-Bessel series and axially independent cylindrical waves:**

The solutions (10) of the wave equation (5) would give rise to an axially independent cylindrical wavelet

$$\mathbf{F} = J_\eta(k\rho)\exp\{(j\omega - (\sigma/2\varepsilon))t + j\eta\phi\} \tag{12}$$

associated with the frequency $\omega$ and the wave number $k$ satisfying the nonlinear relation (11). In order to match the solution (12) on the boundaries $\partial K$ of the model 'K' it is essential to sum up the same solution in the form of Fourier-Bessel series

$$\mathbf{F} = \sum_{i \in J^+} \mathbf{A}_i J_\eta(k_i\rho)\exp\{j\eta\phi + (j\omega - (\sigma/2\varepsilon))t\}. \tag{13}$$

**Determination of Groove fields belonging to the wedge regions $R_i$ ($i = 1, 2$) subject to Dirichlet's boundary conditions on $\partial K$:**

Assuming Dirichlet's conditions

$$\left.\begin{array}{l} F\big|_{OA} = F_1(x',0,0), \quad F\big|_{AC} = F_2(a, y',0) \\ F\big|_{OB} = F_3(0, y',0) \text{ and } F\big|_{BC'} = F_4(x',-b,0) \end{array}\right\} \tag{14}$$

on the faces OA, AC, OB and BC' of the model K, one can arrive at the following pair of dual series relation by matching the Fourier-Bessel series (13) with the function $F_i$ ($i = 1, 2, 3, 4$) given by (14) for $t = 0$

$$\begin{array}{l} \sum_{i \in J^+} \mathbf{A}_i J_\eta(k_i\rho) = F_1(x',0,0)e^{-\eta\phi j} \text{ for } 0 \le \rho \le a, \\ \sum_{i \in J^+} \mathbf{A}_i J_\eta(k_i\rho) = F_2(a, y',0)e^{-\eta\phi j} \text{ for } a \le \rho \le d. \end{array} \tag{15}$$

Now, using the oblique transformation [13]

$$\begin{array}{l} x_1 = \rho\cos\phi = x'\cos\theta_0 - y'\cos(\theta_0 + \beta) \\ x_2 = \rho\sin\phi = -x'\sin\theta_0 + y'\sin(\theta_0 + \beta) \end{array} \tag{16}$$

one can arrive at the coordinates

$$x' = \rho \sin(\theta_0 + \phi + \beta)/\sin \beta,$$
$$y' = \rho \sin(\theta_0 + \phi)/\sin \beta \quad \text{and}$$
$$\rho_{AC} = \rho|_{\text{Face } AC} = a \sin \beta / \sin(\theta_0 + \phi + \beta),$$
$$\rho_{BC'} = \rho|_{\text{Face } BC'} = -b \sin \beta / \sin(\theta_0 + \phi).$$

Hence, one can further express the dual equations in the form

$$\sum_{i \in J^+} \mathbf{A}_i J_\eta(k_i \rho) = f(\rho) \text{ for } 0 \leq \rho \leq d \tag{17}$$

where

$$f(\rho) = e^{\eta \theta j} F_1(\rho, 0, 0) \text{ for } 0 \leq \rho \leq a$$

and

$$e^{-\eta \phi j} F_2(a, y'_{AC}, 0) \text{ for } a \leq \rho \leq d \quad (-\theta \leq \phi \leq 0).$$

Now using Lommel's integral [14] for orthogonality of $J_\eta(k_i \rho)$ in the interval $0 \leq \rho \leq d$ one can express $A_i$ in the form

$$\frac{1}{2} A_i d^2 J_{\eta+1}^2 (dk_i) = \int_{\rho=0}^{d} \rho J_\eta(k_i \rho) f(\rho) d\rho \tag{18}$$

where $k_i$ is the $i^{\text{th}}$ positive zero of $J_\eta(dk) = 0$. Combining (17) and (18) one can further arrive at the result

$$\frac{1}{2} \mathbf{A}_i d^2 J_{\eta+1}^2(dk_i) = e^{\eta \theta j} \int_{\rho=0}^{a} \rho J_\eta(k_i \rho) F_1(\rho, 0, 0) d\rho$$
$$+ \int_{\rho_{AC}=a}^{d} \rho_{AC} e^{-\eta \phi j} J_\eta(k_i \rho_{AC}) F_2(a, y'_{AC}, 0) d\rho_{AC} \tag{19}$$

Now, combining (18) and (19) the unknown coefficients '$A_i$' may be precisely determined by means of the formula

$$\frac{1}{2} A_i d^2 J_{\eta+1}^2(dk_i) A_i = e^{\eta \theta j} \int_{\rho=0}^{a} \rho J_\eta(k_i \rho) F_1(\rho, 0, 0) d\rho + I_1$$

where

$$I_1 = e^{\eta j (\theta + \beta)} (a \sin \beta)^2 \int_{t=\cosec \beta}^{\cosec(\theta + \beta)} t \exp(-\eta j \cosec^{-1} t) J_\eta(k_i a \sin \beta t) F_2(a, y'_{AC}, 0) dt \tag{20}$$

**Cylindrical wave functions:**

The expression (13) represents a cylindrical wave function

$$\mathrm{F}(\rho, \phi, t) = \Phi^{\mathrm{F}}(\rho, \phi) e^{-\sigma t / 2\varepsilon} e^{j \omega t}, \tag{21}$$

where $\Phi^{\mathbf{F}}(\rho,\phi) = \sum_{i \in J^+} \mathbf{A}_i(F) J_\eta(\rho k_i) e^{n\phi j}$ stands for the free space axially independent cylindrical wave associated with the frequency 'ω' and the wave number $k$, satisfying the nonlinear relation (11).

**Determination of current density $J$:**

A current density is constituted by the conduction current $J_c$ and the displacement current $J_d$ according to Maxwell's theorem in electromagnetics and thus one can express $J$ in the form

$$\mathbf{J} = \mathbf{J}_c + \mathbf{J}_d = \sigma \mathbf{E}(\rho,\phi,t) + \varepsilon \frac{\partial \mathbf{E}}{\partial t}(\rho,\phi,t) \qquad (22)$$

Now, combining the relations (21) and (22), $J$ may be finally expressed in the following form

$$\mathbf{J} = \Phi^{\mathbf{E}}(\rho,\phi) e^{t((\sigma/2\varepsilon)-j\omega)}(\sigma/2 + j\omega\varepsilon) \qquad (j=\sqrt{-1}), \qquad (23)$$

where $\Phi^{\mathbf{E}}(\rho,\phi)$ stands for a cylindrical wave function associated with the electric field intensity $E$ which is given by the relations:

$$\mathbf{E}(\rho,\phi,t) = \Phi^{\mathbf{E}}(\rho,\phi) \exp\{(j\omega - (\sigma/2\varepsilon))\}t$$

and

$$\Phi^{\mathbf{E}}(\rho,\phi) = \sum_{i \in J^+} \mathbf{A}_i(\mathbf{E}) J_\eta(\rho k_i) e^{\eta \phi j}.$$

**Determination of Powers of the Groove fields:**

Powers of the groove fields $F_q \in R_q$ $(q=1,2)$ are precisely determined by computing the values of integrals $\int_{R_q} |F_q|^2 dR$ $(q=1,2)$ successively. The present integrals are evaluated over the oblique triangular regions $R_i$ (Fig. 3) by means of the transformations [13]

$$\begin{aligned} x_1 &= \rho\cos\phi = x'\cos\theta_0 - y'\cos(\theta_0 + \beta), \\ x_2 &= \rho\sin\phi = x'\sin\theta_0 + y'\sin(\theta_0 + \beta), \end{aligned} \qquad (24)$$

$$P_{R_q} = \int_{R_q} |\sum_{i \in J^+} A_i J_\eta(\rho k_i) e^{\eta \phi j} G(t)|^2 dR \quad (q=1,2), \qquad (25)$$

by virtue of the expression (13).

Now, using the property of complex numbers one can further express (25) in the form

$$P_{R_q} = \int_{R_q} \left[ \sum_{i \in J^+} A_i^* J_\eta (\rho k_i) e^{-\eta \phi j} \right] \left[ \sum_{i \in J^+} A_i J_\eta (\rho k_i) e^{\eta \phi j} \right] |G(t)|^2 \, dx'dy' \sin \beta, \tag{26}$$

for real values of $\eta, \rho$ and $k_i$.

Apparently, the transformations (24) lead to the area element

$$dx'dy' = \frac{\partial(x', y')}{\partial(\rho, \phi)} d\rho \, d\phi = \rho \, d\rho \, d\phi \tag{27}$$

which on combining with (26) further leads to the results

$$P_{R_1} = \sin \beta \, |G(t)|^2 \int_{\phi=\theta_0}^{0} \int_{\rho=0}^{\rho_{AC}} \left[ \sum_{i \in J^+} |A_i|^2 + 2 \operatorname{Re}(A_i A_j^*) \right] J_\eta^2 (\rho k_i) \rho \, d\rho \, d\phi \tag{28}$$

and

$$P_{R_2} = \sin \beta \, |G(t)|^2 \int_{\phi=\theta_0-\beta}^{\pi} \int_{\rho=0}^{\rho_{BC'}} \left[ \sum_{i \in J^+} |A_i|^2 + 2 \operatorname{Re}(A_i A_j^*) \right] J_\eta^2 (\rho k_i) \rho \, d\rho \, d\phi, \tag{29}$$

where $\rho_{AC} = a \sin \beta / \sin(\theta_0 + \phi + \beta)$ and $\rho_{BC'} = b \sin \beta / \sin(\theta_0 + \phi)$.

Now, evaluating the $'\rho'$ integral in (28) and (29) successively one can arrive at the results

$$P_{R_1} = (a \sin \beta)^{2\eta+2} |G(t)|^2 (\sin \beta) T^*, \tag{30}$$

$$P_{R_2} = (b \sin \beta)^{2\eta+2} |G(t)|^2 (\sin \beta) U^*, \tag{31}$$

where

$$T^* = \sum_{i \in J^+, \eta=0}^{\infty} \frac{(-1)^n (k_i/2)^{2\eta+2m} [|A_i|^2 + 2 \operatorname{Re}(A_i A_j^*)] \Gamma(2\eta + 2m + 1) (a \sin \beta)^{2m}}{m!(2\eta + 2m + 2) \Gamma(2\eta + m + 1) \Gamma(\eta + m + 1)} I_{2\eta+2m+2}, \tag{32}$$

$$I_{2\eta+2m+2} = \int_{\phi=\theta_0}^{0} \operatorname{cosec}^{2\eta+2m+2} (\theta_0 + \phi + \beta) d\phi = \int_{\alpha=0}^{\theta_0} \operatorname{cosec}^{2\eta+2m+2} (\alpha + \beta) d\alpha, \tag{33}$$

$$(2\eta + 2m)(2\eta + 2m + 1) I_{2\eta+2m+2}$$
$$= (2\eta + 2m + 1)[-\operatorname{cosec}^{2\eta+2m}(\theta_0 + \beta) \cot(\theta_0 + \beta) \tag{34}$$
$$+ \operatorname{cosec}^{2\eta+2m} \beta \cot \beta] + 4(\eta + m)^2 I_{2\eta+2m},$$

$$U^* = \sum_{i \in J^+, \eta=0}^{\infty} \frac{(-1)^m (k_i/2)^{2\eta+2m} [|A_i|^2 + 2\operatorname{Re}(A_i A_j^*)] \Gamma(2\eta+2m+1) (b\sin\beta)^{2m}}{m!(2\eta+2m+2) \Gamma(2\eta+m+1) \Gamma(\eta+m+1)} J_{2\eta+2m+2},$$

(35)

$$J_{2\eta+2m+2} = \int_{\phi=(-\theta_0+\beta)}^{-\pi} \cosec^{2\eta+2m+2}(\theta_0+\phi)\,d\phi = \int_{\alpha=-\beta}^{\theta_0-\pi} \cosec^{2\eta+2m+2}\alpha\,d\alpha \tag{36}$$

and

$$\begin{aligned}(2\eta+2m)(2\eta+2m+1)I_{2\eta+2m+2} \\ = (2\eta+2m+1)[-\cosec^{2\eta+2m}(\theta_0-\pi)\cot(\theta_0-\pi) \\ + \cosec^{2\eta+2m}(-\beta)\cot(-\beta)] + 4(\eta+m)^2 I_{2\eta+2m}.\end{aligned} \tag{37}$$

**Conclusions:**

The present paper consists of the results concerning an axially independent EM field associated with an echellete model. The model happens to be a vital part of a periodic echellete antenna forming a corrugated structure. The present field of study happens to be equivalent to EM boundary value problems. Important EM problem due to Dirichlet has been taken into consideration subject to the prescribed values of the said EM field. The wave nature of the present EM field has been justified by arriving at the nonlinear relation $4\mu\varepsilon k^2 = \mu^2\sigma^2 + 4\omega^2\mu^2\varepsilon^2$, where $\mu, \varepsilon, \sigma$ stand for the permeability, permittivity and the conductivity of medium associated with the concerning EM wave and $k, \omega$ stand for the wave number $(k = 2\pi/\lambda)$ and the frequency $(c = \omega\lambda)$ of the said EM wave. Groove fields and their associated powers have been determined which is essential for precise designing of triangular corrugated structures for studying the blazing effect of a propagating EM wave.

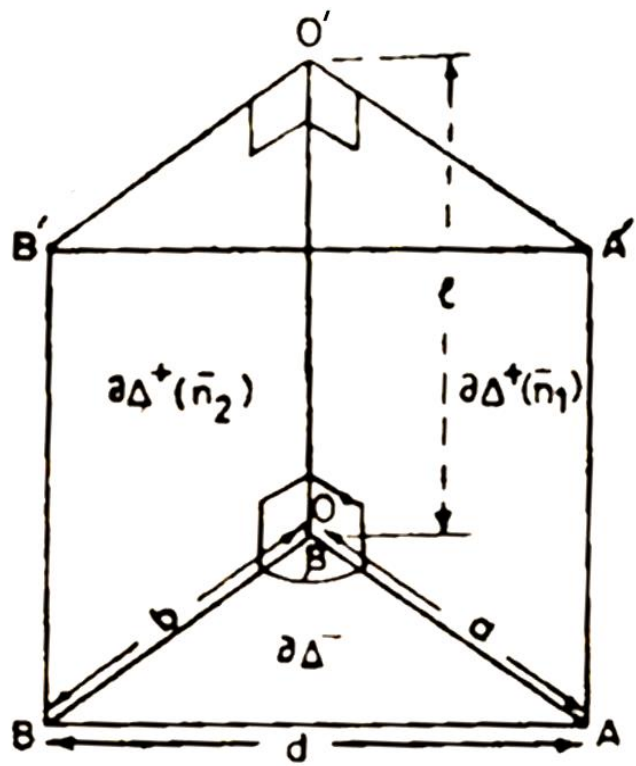

Fig. 1

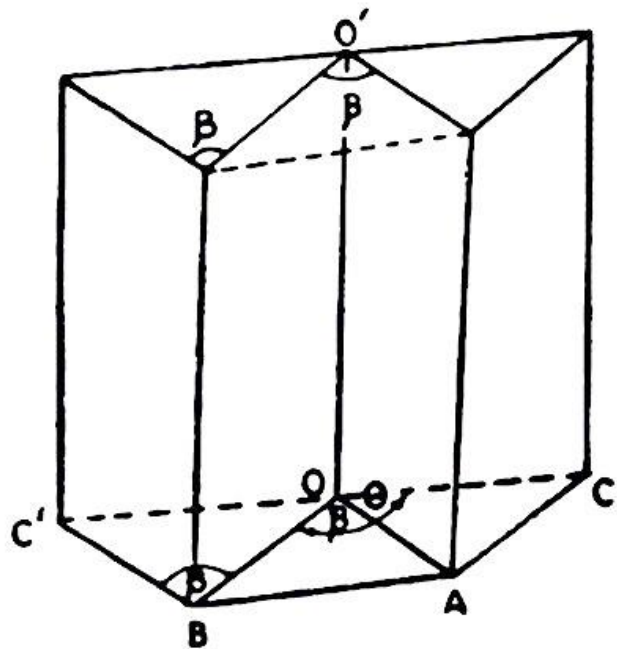

Fig. 2

Fig. 3

**Captions of the Figure**

**Figure 1:**

A convex triangular prism of dimensions a, b, d and with its flare angle 'β'; $OO'$ is perpendicular to the planes $\Delta^s$ OAB and $O'A'B'$.

**Figure 2:**

A model 'K' consists of a triangular prism formed $\Delta^s$ OAB and $O'A'B'$ and its adjacent groove regions formed by the sides $BC'$ and AC and the sides parallel to $OO'$, OA and OB.

**Figure 3:**

    (i)    $\Delta OAB$ is the vertical section of obstacle K.

    (ii)    $\Delta^s$ $OAC$ and $OBC'$ are the vertical sections of the wedge regions $R_q$ $(q=1,2)$.

    (iii)    OA, OB and Ab are parallel to BC', AC and OC' respectively together with OA=BC'=a, OB=AC=b, and AB=OC=d.